\documentclass[12pt]{iopart}
\usepackage{graphicx}
\begin{document}

\title{
Vortex States in Archimedean Tiling Pinning Arrays 
} 
\author{
D. Ray$^{1,2}$, C. Reichhardt$^{1}$, 
and C. J. Olson Reichhardt$^{1}$ 
} 
\address{
$^{1}$ Theoretical Division,
Los Alamos National Laboratory, Los Alamos, New Mexico 87545 USA\\ 
$^{2}$ Department of Physics, University of Notre Dame, Notre Dame,
Indiana 46556
} 
\ead{cjrx@lanl.gov}

\begin{abstract}
We numerically study vortex ordering and pinning in  
Archimedean tiling substrates composed of square and triangular 
plaquettes. The two  
different plaquettes become occupied at different vortex densities, producing
commensurate peaks in the magnetization at non-integer matching fields.
We find that as the field increases, 
in some cases the fraction of occupied pins can decrease 
due to the competition between fillings of the different plaquette types.
We also identify a number of different types of vortex orderings as function of
field at integer and non-integer commensurate fillings.   
\end{abstract}
\maketitle

\section{Introduction}
There have been extensive studies on vortex ordering and pinning 
for superconductors with periodic arrays of pinning sites 
where the arrays have square or triangular order. 
In these systems the critical current or the force needed
to depin the vortices passes through maxima due to commensuration
effects that occur
when the number of vortices is an integer multiple of the 
number of pinning sites 
\cite{1,2,3,4,5,6,7,8}.
The field at which the number of vortices equals the number of pinning sites is 
labeled $B_\phi$,
so that commensurate peaks arise at $B/B_{\phi} = n$, where $n$ is an integer. 
At these matching fields various types of vortex crystalline states can form 
as has been directly observed in experiments and 
confirmed in simulations \cite{9,10,11,12}. It is also possible
for fractional matching commensurability effects to occur 
at fillings of $B/B_{\phi} = n/m$, where $n$ and $m$ are integers. 
For square and triangular arrays, 
these fractional matching peaks are typically smaller
than the integer matching peaks. The
fractional matchings are associated with
different types of ordered or partially ordered vortex arrangements 
\cite{13,14,15,16,17}. 

In studies of rectangular pinning arrays,
a crossover from matching of the full 
two-dimensional (2D) array to matching with only one length scale
of the array   
occurs for increasing field \cite{18,19}. 
Honeycomb and kagome pinning arrays \cite{20,21,22,23,24} 
are constructed by the systematic dilution of a triangular pinning array.    
In a honeycomb array, 
every third pinning site of the triangular array is removed, while for a 
kagome array, every fourth pinning site is removed. 
In these systems there are strong commensurability 
effects at both integer and noninteger matching fields, 
where the noninteger matchings correspond to integer matchings of the 
original undiluted triangular array.  
A similar effect can occur for the random dilution of a
triangular pinning lattice, where commensuration effects 
occur at integer matching fields as well as 
at the noninteger matching fields corresponding 
to the integer matching fields of the original undiluted pinning array 
\cite{25,26}. 
Other periodic pinning array geometries have also been studied  
which have artificial vortex spin ice arrangements \cite{27} 
or composite arrays of smaller and larger 
coexisting pinning sites \cite{28,29}.

\begin{figure}
\includegraphics[width=3.5in]{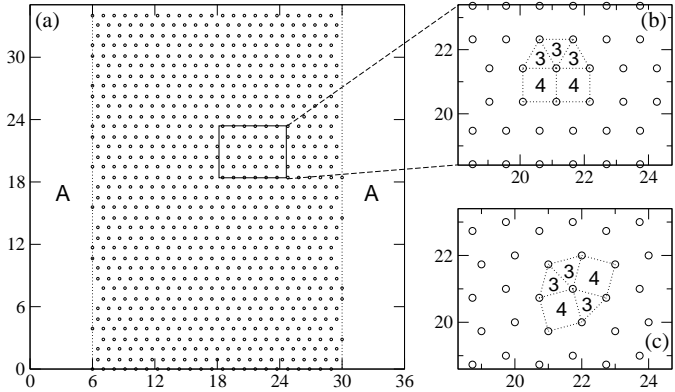}
\caption{ 
(a) The sample geometry shown with the 
$3^34^2$
pinning array. 
Circles represent pinning sites, and vortices are 
added in the unpinned regions marked $A$.
(b,c) Pinning arrays constructed from Archimedean tilings. 
Plaquettes around one pinning site
are highlighted with dotted lines, and labeled 
with the number of sides of the plaquette. The side numbers
read off in a clockwise order around the pinning site 
are used to name the array. (b) The 
$3^34^2$
pinning
array, also called an elongated triangular tiling.
(c) The 
$3^2434$
pinning array, 
also called a snub square tiling. 
}
\label{fig:1}
\end{figure}

\begin{figure}
\includegraphics[width=3.5in]{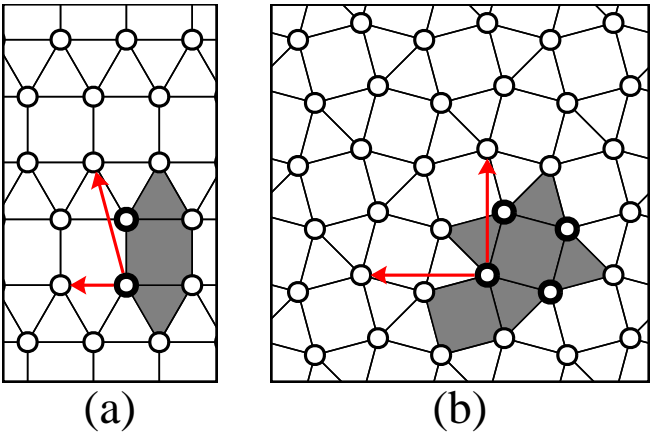}
\caption{ 
Bases for the various arrays. For each array, 
the pinning site basis generating the pinning array is 
plotted with thick dark circles, while
the plaquette basis generating the tiling 
is indicated by the shaded polygons. 
Red arrows show the elementary translation vectors.   
(a) 
$3^34^2$
array. (b) 
$3^2434$ 
array.
}
\label{fig:11}
\end{figure}

Here we propose and study new types of periodic pinning geometries that 
can be constructed from Archimedean tilings of the plane. 
In contrast to a regular tiling where a single type of regular polygon 
(such as a square or equilateral triangle) is used to tile the plane, 
an Archimedean tiling uses two or more different polygon types. 
We consider two examples of 
Archimedean tilings constructed 
with a combination of square and triangular plaquettes,
the elongated triangular tiling illustrated in figure \ref{fig:1}(b), and
the snub square tiling shown in figure \ref{fig:1}(c). 
The plaquettes around one vertex in each tiling are highlighted with
dotted lines and marked with the number of sides. 
The tilings are named by reading off the number of sides in a clockwise
order around the pinning site, giving 33344 
(also written as $3^{3}4^{2}$) 
for the tiling in
figure \ref{fig:1}(b), and 33434 (or $3^{2}434$) for the tiling in 
figure \ref{fig:1}(c). 
In figure \ref{fig:11} one can see the basis of plaquettes 
for each tiling 
which is translated in order to generate the full tiling. 
For each tiling, we place pinning sites 
at the vertices of the polygons to generate a pinning array. 
In figure \ref{fig:1}(a) we illustrate the full simulation geometry 
for the 
$3^34^2$ 
pinning array. 
There are additional types of Archimedean tilings \cite{Grunbaum}, but 
here we concentrate on only the two tilings illustrated in figure \ref{fig:1}; 
pinning arrays constructed from other tilings will be described in a future 
work \cite{unpub}.

It might be expected that the behavior of the 
Archimedean pinning arrays 
would not differ significantly from that of purely square or triangular
pinning arrays;
however, we find that the different plaquette types in the 
Archimedean tilings compete. 
The triangular and square plaquettes comprising the array 
have equal side lengths $a$; 
thus, from simple geometry, 
the distance from the center of a plaquette to any of its vertices 
will be larger for the square plaquette
($a/\sqrt{2}$ versus $a/\sqrt{3}$). 
As a consequence, interstitial vortices prefer to occupy 
square plaquettes rather than triangular plaquettes.
This produces several strong matching effects at 
certain non-integer filling fractions where the vortices are ordered, 
and suppresses commensurability effects at certain integer fillings where
the vortices are disordered. 
In some cases we even find a drop in the pinning site occupancy with
increasing magnetic field.      
We also observe several partially ordered states as well as
different fractional fields and submatching 
fields that do not arise in regular square or periodic pinning arrays.

\section{Simulation and System}   
In this work we utilize flux gradient density simulations 
as previously employed to study 
vortex pinning in random \cite{37}, periodic \cite{38}, 
and conformal pinning arrays \cite{30}.
The sample geometry is illustrated in figure \ref{fig:1}(a). 
We consider a 2D system with
periodic boundary conditions in the $x$- and $y$-directions. 
The sample size 
$36\lambda\times 36\lambda$ is measured in units
of the penetration depth $\lambda$. 
Our previous
studies 
indicate that this size of simulation box is sufficiently large to obtain
experimentally relevant magnetization curves.
The system represents a 2D slice
of a three-dimensional 
type-II superconductor 
with a magnetic field applied in the 
perpendicular (${\hat z}$) direction, 
and we assume that the vortices behave as rigid objects. 
We work in the London limit of vortices with pointlike cores,  
where the coherence length $\xi$ is much smaller than $\lambda$. 
The pinning sites are located in a $24\lambda$ wide region in the middle
portion of the sample, with pin-free regions labeled $A$ 
in figure \ref{fig:1}(a) on
either side.
Vortices are added to region $A$ during the simulation, and their density
in this region
represents the externally applied field $H$. The vortices enter the pinned
region from the edges and form a Bean gradient \cite{Bean}. 
The motion of the vortices   
is obtained by integrating the following overdamped equation of motion:
\begin{equation}  
\eta \frac{d {\bf R}_{i}}{dt} = 
 {\bf F}^{vv}_{i} + {\bf F}^{vp}_{i}. 
\end{equation} 
Here $\eta$ is the damping constant
which is set equal to 1,  
${\bf R}_{i}$ is the location of vortex $i$,  
${\bf F}^{vv}_{i}$ is the vortex-vortex interaction force,
and ${\bf F}^{vp}_{i}$ is the force from the pinning sites.   
The vortex-vortex 
interaction force is 
given by 
${\bf F}^{vv}_{i} = \sum_{j\neq i}F_{0}K_{1}(R_{ij}/\lambda){\hat {\bf R}_{ij}}$,
where $K_{1}$ is the modified Bessel function, 
$R_{ij} = |{\bf R}_{i} - {\bf R}_{j}|$,
$ {\hat {\bf R}_{ij}} = ({\bf R}_{i} - {\bf R}_{j})/R_{ij}$, 
and $F_{0} = \phi^{2}_{0}/(2\pi\mu_{0}\lambda^3)$, 
where $\phi_{0}$ is the flux quantum and $\mu_{0}$ is the permittivity.    
The pinning sites are modeled as $N_{p}$ non-overlapping parabolic traps with
${\bf F}^{vp}_{i} = \sum^{N_{p}}_{k= 1}(F_{p}R^{(p)}_{ik}/r_{p})\Theta((r_{p}  
-R^{(p)}_{ik})/\lambda){\hat {\bf R}^{(p)}}_{ik}$,  
where $\Theta$ is the Heaviside step function, 
$r_{p} = 0.12\lambda$ is the pinning radius, $F_{p}$ is the pinning strength, 
${\bf R}_k^{(p)}$ is the location of pinning site $k$,
$R_{ik}^{(p)} = |{\bf R}_{i} - {\bf R}_{k}^{(p)}|$, and
$ {\hat {\bf R}_{ik}^{(p)}} = ({\bf R}_{i} - {\bf R}_{k}^{(p)})/R_{ik}^{(p)}$. 
We consider three pinning densities of $n_{p} = 1.0/\lambda^2$, 
$2.0/\lambda^{2}$, and $0.5/\lambda^{2}$, as well as a range of $F_{p}$ values.
All forces are measured in units of $F_{0}$ and lengths in units of $\lambda$. 
The magnetic field $H$ is measured in terms of the matching field $H_{\phi}$ 
where the density of vortices equals the density of pinning sites.  
We measure only the first quarter of the magnetic hysteresis loop, which is
sufficient to identify the different commensuration effects.
We concentrate on the regime where only one vortex is trapped per 
pinning site, so that for fields greater than the first matching 
field, the additional vortices are located in the interstitial regions.  

\begin{figure}
\includegraphics[width=3.5in]{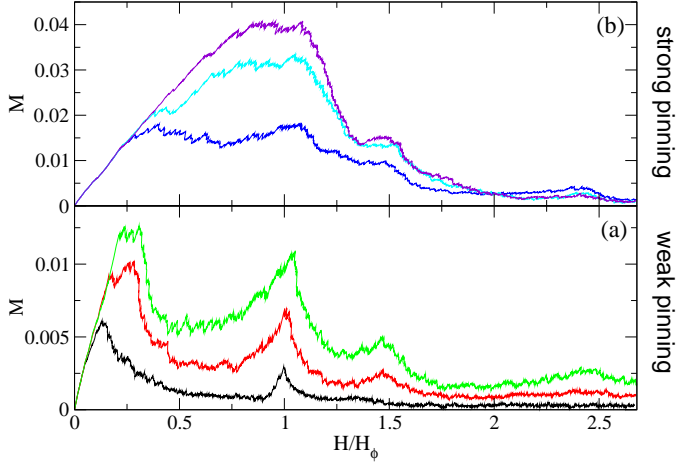}
\caption{ 
The magnetization $M$ vs $H/H_{\phi}$ for the 
$3^34^2$ 
lattice 
from figure \ref{fig:1}(b) 
with $n_{p} = 1.0/\lambda^2$. 
The various curves correspond to different values of the 
pinning strength $F_p$, which increases from bottom to top.  
(a) $F_{p} = 0.1$ (black), $0.2$ (red), 
$0.3$ (green). (b) $F_p=0.5$ (blue), 
$0.8$ (cyan), and $1.0$ (violet).  
}
\label{fig:2}
\end{figure}

\section{Elongated Triangular Tiling 
($3^34^2$)
Lattice}
We first consider the elongated triangular tiling or 
$3^34^2$
lattice shown in figure \ref{fig:1}(b),
which consists of rows of square plaquettes 
separated by rows of triangular plaquettes.  
In figure \ref{fig:2} we plot the magnetization $M$ vs $H/H_{\phi}$, 
where $M = (1/4\pi V) \int (H - B) dV$ 
with $B$ representing the field inside the pinned region 
and $V$ its area.
Here we use a pinning density of $n_{p} = 1.0/\lambda^2$. 
The critical current is proportional to the width of the 
magnetization loop, so a peak in $M$ corresponds to a peak 
in the critical current.   
figure \ref{fig:2}(a) shows $M$ for $F_{p} = 0.1$, $0.2$, and $0.3$.
In each case there is a peak in $M$ associated with the
first matching condition of $H/H_{\phi} = 1.0$; 
however, at $H/H_{\phi} = 2.0$ there is no peak. 
Instead, peaks appear at $H/H_{\phi} = 1.5$ and $2.5$.  
In figure \ref{fig:2}(b) we show $M$ versus $H/H_\phi$ for samples with stronger
pinning of
$F_{p} = 0.5$, $0.8$, and $1.0$. Here 
the peak at $H/H_{\phi} = 1.0$ is obscured by
the initial rise in the magnetization; however, peaks are still 
present at  $H/H_{\phi} = 1.5$ and $2.5$. 

\begin{figure}
\includegraphics[width=3.5in]{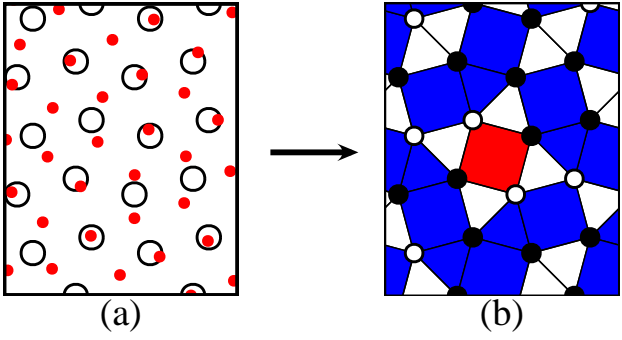}
\caption{
(a) Part of a sample vortex configuration in a 
$3^2434$
array. 
Large open circles: pinning sites; small filled circles: vortices. 
(b) Corresponding plaquette occupancy diagram.
Filled black circles denote occupied pinning sites, 
open circles denote empty pinning sites, and colored tiles indicate
plaquettes occupied by one (blue) or more (red) 
interstitial vortices.  
}
\label{fig:3}
\end{figure}

\subsection{States at and above first matching field}
In order to understand better the vortex states at fields beyond the
first matching peak for both types of pinning arrays, we  
analyze the plaquette occupancy using the
tiling coloring scheme illustrated in figure \ref{fig:3}. 
In figure \ref{fig:3}(a), we show a sample configuration of 
pinning sites and vortices for the 
$3^2434$ 
lattice from Fig.~\ref{fig:1}(c). 
Figure \ref{fig:3}(b) shows the same configuration represented as a 
plaquette occupancy diagram, 
with pinning sites marked either
dark or light depending on whether they are filled or empty, and plaquettes
marked either with white fill, dark fill, or light fill depending on whether
they are occupied by zero, one, or more than one 
interstitial vortex, respectively.

\begin{figure}
\includegraphics[width=3.5in]{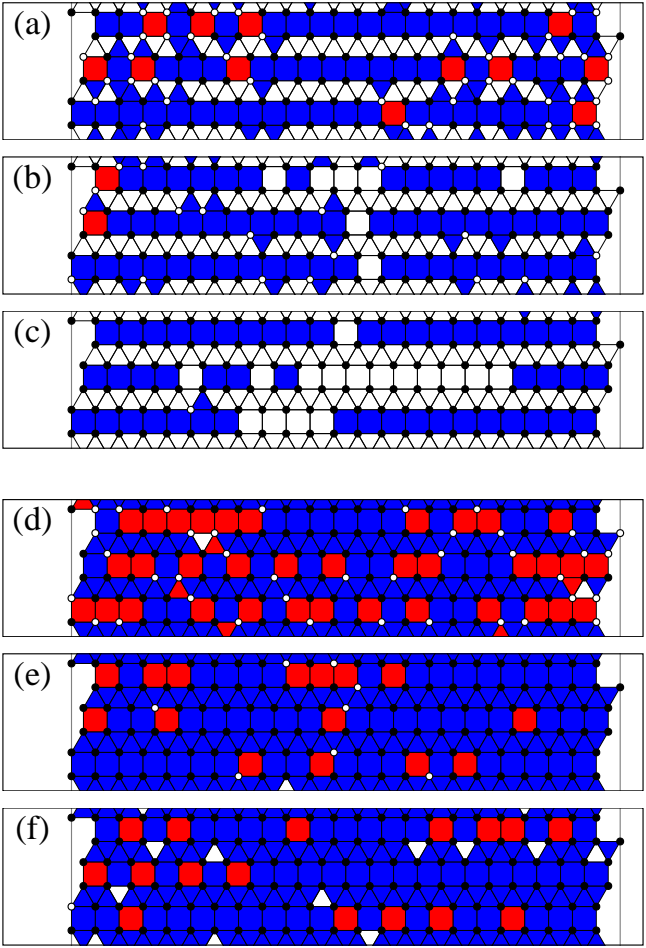}
\caption{
Plaquette occupancy in representative strips of the 
$3^34^2$ 
sample from figure \ref{fig:1}(b),
at field levels corresponding to peaks in figure \ref{fig:2}. 
The full width of the pinned region is shown.
The coloring scheme is described in figure \ref{fig:3}.
(a) $H/H_{\phi}=1.55$, $F_p=0.2$.
(b) $H/H_{\phi}=1.55$, $F_p=0.5$.
(c) $H/H_{\phi}=1.55$, $F_p=1.0$.
At this field the square plaquettes are predominantly filled.
(d) $H/H_{\phi}=2.56$, $F_p=0.2$.
(e) $H/H_{\phi}=2.56$, $F_p=0.5$.
(f) $H/H_{\phi}=2.56$, $F_p=1.0$.
At this field most plaquettes are filled. 
}
\label{fig:4}
\end{figure}

We focus on the field values at which peaks
in $M$ appear 
in figure \ref{fig:2}. 
At $H/H_{\phi} = 1.0$,  
each pinning site captures one vortex,  
while at $H/H_{\phi} = 1.5$,  
the additional vortices predominantly 
occupy the interstitial regions at the center of the
square plaquettes. 
This is shown in figure \ref{fig:4}(a-c) 
where we plot the plaquette occupancy at $H/H_{\phi} = 1.55$ 
for increasing $F_{p}$. 
(We select a value of $H/H_{\phi}$ slightly higher than
$H/H_{\phi} = 1.5$ to compensate for the Bean gradient since the
field outside the pinning region is larger than the field inside the 
pinned part of the sample.)
As $F_{p}$ is increased, the ideal 1.5 state forms more cleanly.  
For the weak pinning situation 
$F_{p} = 0.2$ in figure \ref{fig:4}(a), there are a handful of unoccupied pinning
sites, and 
interstitial vortices occupy some triangular plaquettes 
and doubly occupy some of the square plaquettes.
At $F_{p} = 0.5$ in figure \ref{fig:4}(b), more of the pinning sites
are filled and the interstitial 
vortices increasingly occupy only the square plaquettes. 
Finally, at strong pinning $F_{p} = 1.0$ in figure \ref{fig:4}(c), 
nearly all pinning sites are occupied, 
since those
vortices which were pinned at the first matching field never depin, 
so that as the field is raised above the first matching field,
interstitial vortices enter the sample along the rows of square plaquettes, 
filling them.
We note that there are a larger number of empty plaquettes at the center 
of the sample for the stronger pinning due to the Bean gradient, 
which is created by the pinning.   

In figure \ref{fig:4}(d,e,f) we show the plaquette fillings at
$H/H_{\phi} = 2.56$ 
corresponding to the final peak in figure \ref{fig:2} 
for $F_{p}= 0.2$, $0.5$, and $1.0$, respectively.
For the highest $F_{p}=1.0$ in figure \ref{fig:4}(f), 
all the pinning sites capture one 
vortex and most of the plaquettes are filled;
however, in a number of locations
there is an empty triangular plaquette accompanied by a doubly
occupied square plaquette.
At the weaker pinning strength 
$F_p=0.2$ in figure \ref{fig:4}(d), essentially all of the plaquettes, 
including the triangular ones, are filled; 
moreover, the weaker pinning leads to a number of unoccupied pinning 
sites.  In these cases the depinned vortices prefer to sit in the square 
plaquettes, making them doubly occupied. 

\begin{figure}
\includegraphics[width=3.5in]{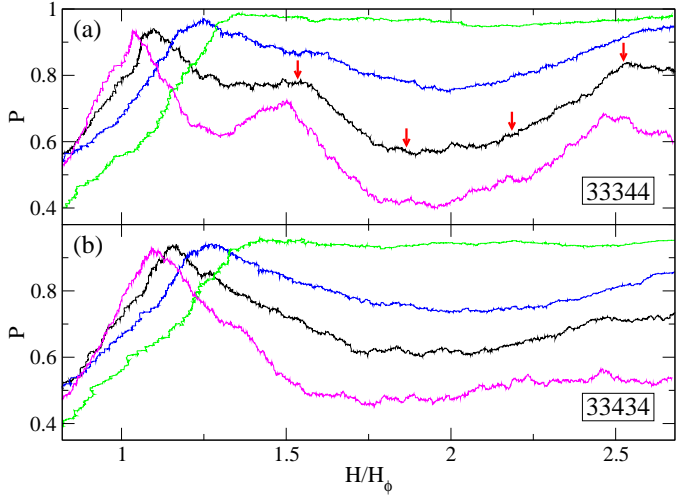}
\caption{
$P$, the fraction of occupied pins, vs $H/H_{\phi}$. 
(a) The 
$3^34^2$ 
array from figure \ref{fig:2}
at $F_{p} = 0.2$, $0.3$, $0.5$, and $0.8$, from bottom right to top right.  
Arrows pointing to various locations on the $F_p=0.3$ curve 
indicate field levels where we illustrate the real-space 
vortex configurations in figure \ref{fig:6}. 
(b) The 
$3^2434$ 
array at 
$F_p=0.2$, $0.3$, $0.5$, and $0.8$, from bottom right to top right. 
}
\label{fig:5}
\end{figure}

At $H/H_{\phi} = 2.0$ the system could in principle form an ordered 
state where every square plaquette is occupied while every other
triangular plaquette is unoccupied; however, we do not observe 
such an ordered state. 
Instead, we find that
as the field is increased from the ordered $H/H_{\phi} = 1.5$ state 
where only the 
square plaquettes are occupied, 
some pinned vortices
are pushed out of the pinning sites when additional vortices try to 
occupy the triangular plaquettes. 
As a result, the pin occupancy actually decreases with increasing field,
as shown in figure \ref{fig:5}(a) where we plot the fraction of occupied pinning 
sites $P$ versus $H/H_{\phi}$ for the 
$3^34^2$
system at four different
values of $F_p$.
There is a peak in $P$ just above 
$H/H_{\phi} = 1.0$ corresponding to the first matching field
where most of the pinning sites are occupied. 
For weaker pinning, $F_{p} \leq 0.5$, 
$P$ declines from its peak but stabilizes 
near $H/H_{\phi} = 1.5$, as illustrated for 
$F_p=0.3$ 
in figure \ref{fig:5}(a).
As $H/H_{\phi}$ increases further, $P$ 
begins to fall substantially when  
vortices start to push their way into 
triangular plaquettes, causing the vortices at the neighboring 
pinning sites to depin.  Near $H/H_{\phi}=2.0$, $P$ passes through a
minimum;
the overall vortex configuration at $H/H_{\phi} = 2.0$ is disordered and 
there is no peak in $M$ at this field.
$P$ then recovers and increases up to 
the ordered state at $H/H_{\phi} = 2.5$, where 
every square and every triangular plaquette can contain 
an interstitial vortex as illustrated in figure \ref{fig:4}(d,e). 
The field level for this state can 
be understood from figure \ref{fig:11}(a), where we see that a basis 
for the 
$3^34^2$ 
tiling has 2 pinning sites, 2 triangular plaquettes, and
1 square plaquette; if all these are singly occupied, we obtain 
a field level of $5/2 = 2.5$. 

\begin{figure}
\includegraphics[width=3.5in]{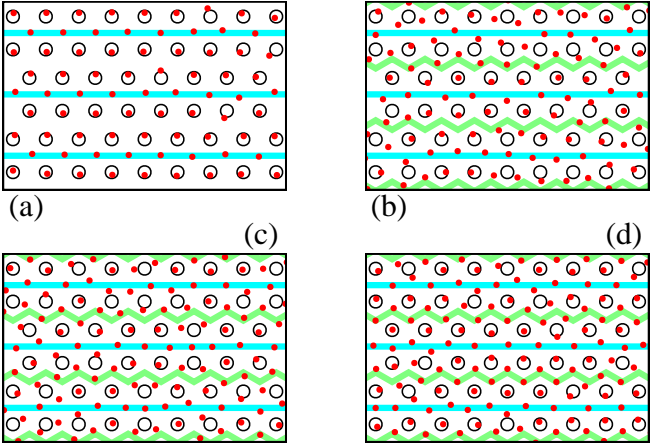}
\caption{
Real space images of vortex configurations 
for the 
$3^34^2$ 
array at $F_p=0.3$ 
as the field $H/H_{\phi}$ is increased from $1.5$ to $2.5$,
showing lines of interstitial vortices buckling and then reforming. 
Open circles: pinning sites; filled dots: vortices; thick lines: guides to
the eye running along rows of plaquettes and connecting their centers,  
which are preferred locations for interstitial vortices when all 
surrounding pinning sites are occupied. 
(a) $H/H_{\phi} = 1.535$; (b) $1.865$; (c) $2.185$; (d) $2.524$.
The field levels shown correspond to the red arrows in figure \ref{fig:5}(a).
}
\label{fig:6}
\end{figure}

It is instructive to visualize the transition from $H/H_{\phi} = 1.5$ to
$H/H_{\phi} = 2.5$ in real space. After the first matching field is 
reached and the pinning sites become occupied, interstitial vortices 
enter the sample between 
the square tiles in neat horizontal lines. This continues until every 
square tile in a row contains one interstitial vortex, giving the 
$H/H_{\phi}=1.5$ 
state shown in figure \ref{fig:6}(a). 
When further vortices attempt to enter, the 
lines of interstitials buckle, as shown in figure \ref{fig:6}(b,c); 
in this process, 
many pinned vortices are dislodged, and the 
triangular plaquettes become
filled in an irregular 
manner. As the sample approaches the ordered 
$H/H_{\phi}=2.5$ state, the lines of interstitial vortices 
re-form along the square plaquettes,
and the triangular 
plaquettes now also fill with lines of interstitials 
which zig-zag due to the alternating orientation of the triangles. 
This is illustrated in figure \ref{fig:6}(d). 

For stronger pinning $F_{p} > 0.5$, 
the pinned vortices do not depin when the
triangular plaquettes start to become occupied around $H/H_{\phi}=2.0$, 
as shown for $F_p=0.8$ in figure \ref{fig:5}(a); 
thus, the buckling phenomenon observed for weaker pinning 
does not 
occur in this case. 
However, the vortex configurations around $H/H_{\phi}=2.0$ are still disordered
even for strong pinning,  
since the triangles do not 
become occupied in an orderly 
manner.

\begin{figure}
\includegraphics[width=3.5in]{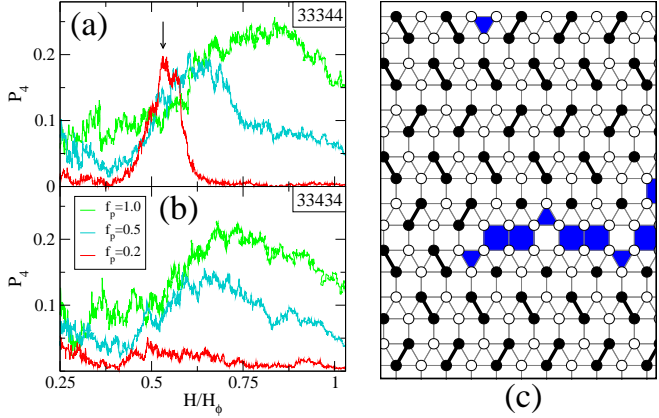}
\caption{
(a) $P_4$, the fraction of vortices with a coordination number of 4, vs
$H/H_{\phi}$ 
for the 
$3^34^2$ 
array in a system with 
$n_{p} = 2.0/\lambda^2$ and 
$F_{p} = 0.2$ (blue), $0.5$ (red) and $1.0$ 
(green). 
Here a peak (marked with an arrow) occurs near $H/H_{\phi} = 0.5$ 
for the weakest pinning. 
(b) $P_4$ vs $H/H_{\phi}$
for the 
$3^2434$ 
array 
at $n_p=2.0/\lambda^2$ and
$F_{p} = 0.2$ (blue), $0.5$ (red) and $1.0$ 
(green); 
the peak from panel (a) is absent. 
(c) A subsection of the $F_{p} = 0.2$ system 
in the 
$3^34^2$ 
sample from panel (a) at $H/H_{\phi}=0.5$. 
Empty circles: unoccupied pinning sites; filled circles: occupied pinning
sites; white squares and triangles: unoccupied square and triangular
plaquettes; filled squares and triangles: square and triangular plaquettes
that each contain one vortex.
A thick heavy line is drawn between pairs of occupied pinning sites to
highlight the herringbone ordering.
}
\label{fig:7}
\end{figure}

\subsection{Submatching states}

We also find 
vortex ordering at some submatching 
fields for the 
$3^34^2$ 
pinning array.
This is more clearly visible in an array with 
higher density 
$n_{p} = 2.0/\lambda^2$ and small $F_{p}$. 
In figure \ref{fig:7}(a) we plot $P_4$, 
the fraction of vortices with a coordination
number of four, versus $H/H_{\phi}$ for samples with 
$3^34^2$ 
pinning arrays
with $F_p=1.0$, 0.5, and 0.2.
We obtain the coordination number $z_i$ of each vortex using a Voronoi
construction of the vortex positions, and take
$P_4=N_v^{-1}\sum_{i=1}^{N_v}\delta(4-z_i)$.
For $F_{p} = 0.2$ there is a peak 
in $P_4$ just above $H/H_{p}  = 0.5$, 
marked with an arrow in figure \ref{fig:7}(a), which corresponds to 
an ordered vortex sub-matching configuration.  
This configuration is illustrated in figure \ref{fig:7}(c), 
where we show the locations
of the vortices and pinning sites and add
a thick line between pairs of occupied pinning sites 
to make the 
ordering more visible.  Here, every other
pinning site captures a vortex and there is an effective dimerization of
the occupied pinning sites along the edges of the triangular plaquettes.
The dimers are tilted $30^{\circ}$ from the $y$ axis in one row and
$-30^{\circ}$ from the $y$ axis in the next row, giving a herringbone
ordering of the type previously studied for dimer molecules adsorbed on
triangular substrates \cite{39}.
As the pinning strength increases, the dimer ordering breaks apart.
For the 
$3^2434$ 
array there is no herringbone ordering, 
as shown by the absence of a peak in $P_4$ in figure \ref{fig:7}(b).  

\begin{figure}
\includegraphics[width=3.5in]{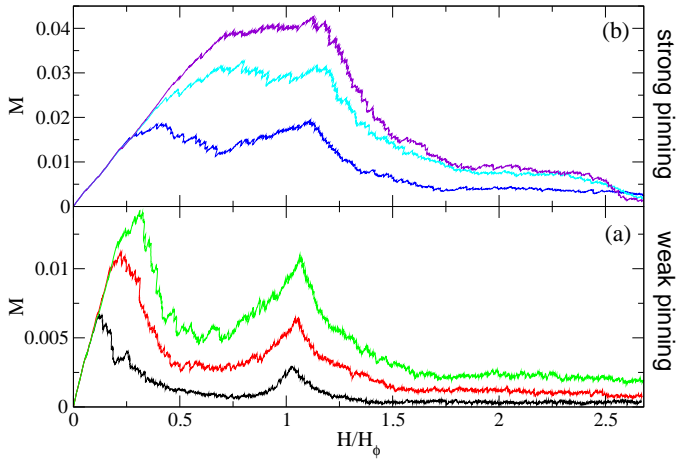}
\caption{ $M$ vs $H/H_{\phi}$ for the 
$3^2434$ 
array 
from figure \ref{fig:1}(c)
for $n_p=1.0/\lambda^2$.
(a) $F_{p} = 0.1$ (black), 0.2 (red), and 
0.3 (green). 
(b) The same for $F_{p} = 0.5$ (blue), 0.8 (cyan) 
and 1.0 (violet).  
}
\label{fig:8}
\end{figure}

\section{Snub Square 
($3^2434$) 
Tiling}  
We next consider the the snub square tilling or 
$3^2434$ 
pinning array 
illustrated in figure \ref{fig:1}(c).
In figure \ref{fig:8}(a) we plot $M$ vs $H/H_{\phi}$ for this array with
$F_{p} = 0.1$, $0.2$, and $0.3$, while samples with
$F_p=0.5$, $0.8$, and $1.0$ are shown in figure \ref{fig:8}(b). 
Here a matching peak occurs at $H/H_{\phi} = 1.0$ but there are no other 
clearly defined peaks at the higher fillings.
For the stronger pinning sample with $F_p=1.0$, the first matching 
peak is obscured by the initial rise of $M$ as shown in figure \ref{fig:8}(b). 
Since there are few additional features in $M$, we
use alternative measurements to show that a variety of partially 
ordered vortex states can occur in this system. 
In particular, we consider the fraction of $n$-fold coordinated
vortices $P_n$, with $n=4$, $5$, $6$, $7$, and $8$, defined as written
earlier for $P_4$.

\begin{figure}
\includegraphics[width=3.5in]{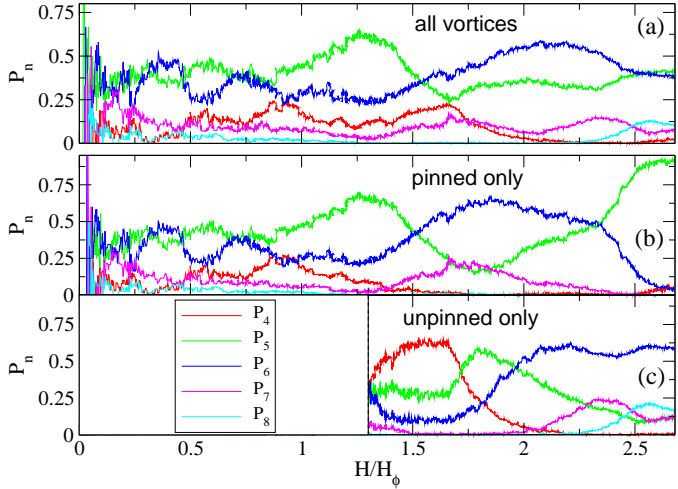}
\caption{
$P_n$ vs $H/H_{\phi}$ for the 
$3^2434$ 
array at $F_{p} =1.0$, with 
$P_{4}$ (red), $P_{5}$ (green), $P_{6}$ (blue),
$P_{7}$ (violet) and $P_{8}$ (cyan). 
(a) All vortices in the system. (b) Pinned vortices only. 
(c) Unpinned vortices only.  
Below $H/H_{\phi}=1.25$ (shaded region in panel (c)),  
there are too few unpinned vortices
to give a clear signal.
}
\label{fig:9}
\end{figure}

\begin{figure}
\includegraphics[width=3.5in]{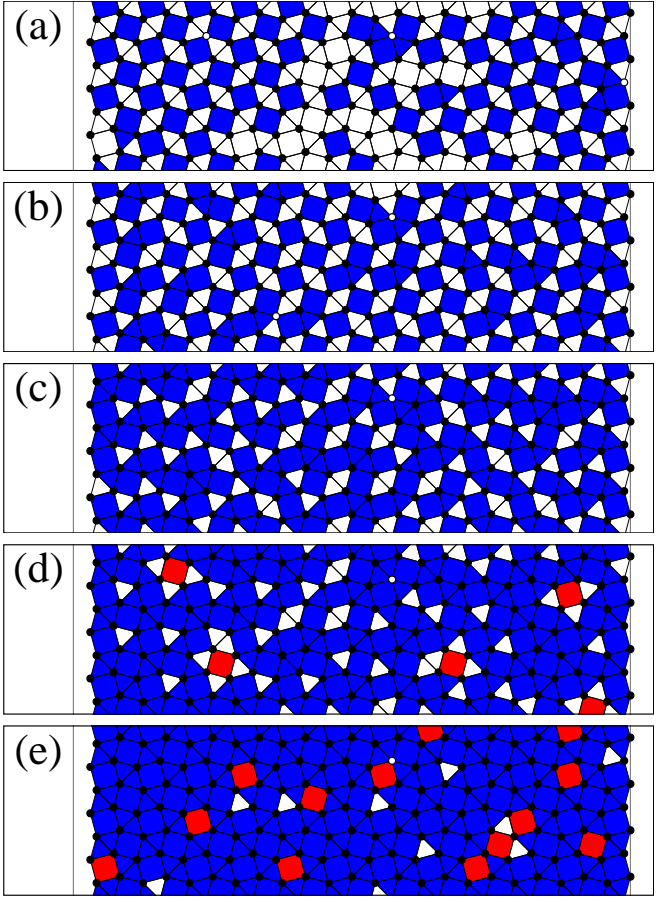}
\caption{
Plaquette occupancy in representative strips of the samples for the
$3^2434$
array with $F_p=1.0$,
at field levels corresponding to peaks in figure \ref{fig:9}(c).  
The full width of the pinned region is shown.
The coloring scheme is described in figure \ref{fig:3}.
(a) $H/H_{\phi} = 1.65$, (b) $1.79$, (c) $2.08$, (d) $2.33$, and (e) $2.56$. 
}
\label{fig:10}
\end{figure}

In figure \ref{fig:9}(a) we plot 
$P_{n}$ with $n=4$ through 8 versus $H/H_{\phi}$ 
for all the vortices in a 
$3^2434$ 
sample with 
strong pinning $F_{p} = 1.0$. 
The ordered states in this system can be more easily distinguished by
separately measuring $P_n$ 
for the pinned vortices only, shown in figure \ref{fig:9}(b), and 
for the interstitial or unpinned vortices only, shown in figure \ref{fig:9}(c).
In particular, 
in figure \ref{fig:9}(c) 
we find 
successive peaks in $P_{4,5,6,7,8}$ for the interstitial vortices. 
Since there are few to no interstitial vortices for $H/H_{\phi}<1.25$, 
we do not show $P_n$ for this field range in figure \ref{fig:9}(c).
The first peak in $P_n$
for the interstitial vortices 
appears in $P_4$ 
just above $H/H_{\phi} = 1.5$ in figure \ref{fig:9}(c). 
Here, most of the pinning sites are occupied 
and the interstitial vortices primarily sit in the square plaquettes,
as indicated in the plaquette occupancy 
plot in figure \ref{fig:10}(a) for $H/H_{\phi} = 1.65$. 
An interstitial vortex in a square plaquette has four nearest neighbors, the
pinned vortices at the edges of the plaquette.
As the field increases, interstitial
vortices begin to occupy isolated triangular plaquettes, 
providing an additional nearest neighbor for the interstitial vortices in
the nearby square plaquettes.
This produces a peak in 
$P_{5}$ at $H/H_{\phi} = 1.79$ in figure \ref{fig:9}(c), 
where nearly all of the filled square plaquettes have one neighboring 
filled triangular plaquette as shown in figure \ref{fig:10}(b).  
At $H/H_{\phi} = 2.08$, there is a peak in $P_{6}$ in figure \ref{fig:9}(c), 
and the corresponding configuration in figure \ref{fig:10}(c) 
shows that many of the square plaquettes now have two neighboring filled
triangular plaquettes.
A peak in $P_7$ appears
at $H/H_{\phi} = 2.33$ in figure \ref{fig:9}(c), and the 
configuration in figure \ref{fig:10}(d) has many square plaquettes 
with three neighboring 
filled triangular plaquettes, as well as a few doubly occupied square 
plaquettes.
Finally, at $H/H_{\phi} = 2.56$ there is a peak in $P_{8}$ 
in figure \ref{fig:9}(c).
The corresponding configurations in figure \ref{fig:10}(e) 
indicate that most of the plaquettes are now filled, with some empty 
triangular plaquettes and some doubly occupied square plaquettes. 

We can understand the field levels at which the peaks in $P_n$ occur by 
considering
figure \ref{fig:11}(b), where we illustrate both the pinning site basis 
and the plaquette basis
which generate the 
$3^2434$
tiling. 
In a ground state configuration at $H/H_{\phi} = 1.0$, there are 
four vortices occupying the four pins comprising the basis, 
with no interstitial vortices.
For the state where each square plaquette 
is occupied by one interstitial vortex, 
there are a total of 6 vortices per basis compared to 4 pins; 
so the field for this state is $H/H_{\phi} = 6/4 = 1.5$. 
Each of the interstitial vortices sitting in a square plaquette has 
4 nearest neighbors, 
the pinned vortices at the corners of the square.  
As $H/H_{\phi}$ increases further, the triangular plaquettes become
occupied one by one,
with each new triangular plaquette interstitial providing an additional 
nearest neighbor for a nearby square plaquette interstitial vortex.
This process gives a total of $7$, 8, 9, or 10 vortices per basis 
leading to fields of 
$H/H_{\phi} = 1.75$, 2.0, 2.25, and $2.5$, respectively. 
Thus, the peaks 
in $P_4$, $P_5$, $P_6$, $P_7$, and $P_8$ 
for the unpinned vortices in figure \ref{fig:9}(c) 
arise simply from the coordination numbers of the interstitial vortices  
occupying the square plaquettes. 
The actual peaks from the simulation 
shown in figure \ref{fig:9}(c) occur at 
$H/H_{\phi} = 1.65$, $1.79$, $2.08$, $2.33$, and $2.56$, 
which are close to the ideal field values.
The small shift in the actual values is 
due to the flux gradient and also to the occasional 
double occupancy of square plaquettes found
in figure \ref{fig:10}(d,e). 
The peaks in $P_{n}$ for the pinned vortices 
shown in figure \ref{fig:9}(b) 
follow immediately from the behavior of the interstitial vortices
described above.
We note in particular that $P_{5}$ becomes nearly
one at $H/H_{\phi} = 2.5$, since almost all the triangular and square 
plaquettes are occupied as shown in figure \ref{fig:10}(e).  
Since each pinning site is surrounded by $5$ 
plaquettes, the nearest neighbors of each pinned vortex 
are surrounded by $5$ interstitial vortices at this filling.

\section{Smaller Pinning Densities}

\begin{figure}
\includegraphics[width=3.5in]{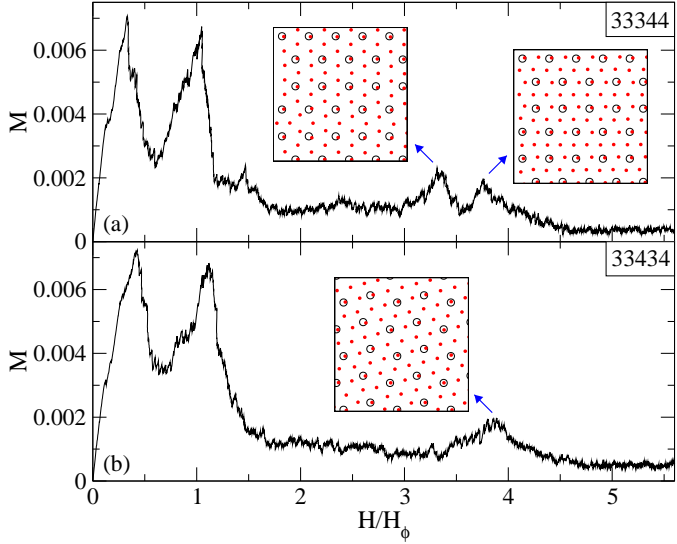}
\caption{
$M$ vs $H/H_{\phi}$ for samples with 
$n_p=0.5/\lambda^2$ and
$F_{p} = 0.3$. 
(a) In the 
$3^34^2$ 
array, there are additional peaks at 
$H/H_{\phi} = 3.5$ and $4.0$.  Insets:
the ordered vortex arrangements that appear at these additional
peaks, indicated by arrows.  In the insets, the open circles are
pinning sites and the filled circles are vortices.
(b) In the 
$3^2434$
array, there is a peak at $H/H_{\phi} = 3.75$; the
inset shows the ordered vortex arrangement that forms
at this field.  In the inset, the open circles are pinning sites
and the filled circles are vortices.
}
\label{fig:12}
\end{figure}

For the case of a pinning density of $n_{p} = 0.5/\lambda^2$ and 
smaller $F_{p}$ we can readily
observe matching peaks up to $H/H_{\phi} = 5.0$. 
In figure \ref{fig:12}(a) we plot $M$ vs $H/H_{\phi}$ 
for the 
$3^34^2$ 
array at $F_{p} = 0.3$ 
and in figure \ref{fig:12}(b) we show the same quantity for 
the 
$3^2434$ 
pinning array. 
In the 
$3^34^2$ 
array, we find the same peaks shown in figure \ref{fig:2} at
$H/H_{\phi} = 1.0$, $1.5$, and $2.5$, and we observe
additional peaks just below 
$H/H_{\phi} = 3.5$ 
and $4.0$ 
In the insets of figure \ref{fig:12}(a) 
we plot the vortex and pinning site locations
at the two new peaks at 
$H/H_{\phi}=3.5$ and $4.0$.
Here, all of the pinning sites are occupied and 
an ordered crystalline arrangement of vortices occurs. 
By directly counting the number of vortices per pinning site in the ordered
part of the sample, we confirm that these are the $3.5$ and $4.0$ fillings.
For the 
$3^2434$ 
array, figure \ref{fig:12}(b) shows a peak in $M$ 
for $H/H_{\phi} = 3.75$ 
corresponding to the ordered vortex array illustrated in the inset.

\section{Summary}

We have investigated ordering and pinning of vortices interacting with 
Archimedean pinning arrays.  The arrays are constructed using
the vertices of Archimedean tilings of squares and triangles, 
and we specifically examine the elongated triangular tiling and
the snub square tiling. 
For the elongated triangular or 
$3^34^2$
array,
we find that beyond the first matching field,
the interstitial vortices first fill the square plaquettes
and subsequently fill the triangular plaquettes, producing
pronounced peaks in the magnetization at noninteger matching fields 
along with an absence of peaks at certain higher integer matching
fields. 
The competition between filling the more confined triangular plaquettes with
single vortices or doubly occupying the larger square plaquettes
can lead
to a decrease in the fraction of occupied pins as the field is increased,
and can produce disordered vortex states at certain integer matching fields.
We also find novel vortex orderings at submatching
fillings, such as herringbone configurations for weak pinning. 
For the snub square or 
$3^2434$
array above the first matching field,
we find that by analyzing the plaquette fillings 
we can correctly predict the appearance of a sequence of
partially ordered
states, 
where interstitial vortices first occupy square plaquettes 
and then fill the triangular plaquettes.   
At higher fields we observe additional commensuration effects at 
noninteger matching fields which correspond to ordered vortex structures. 
Our results can be tested for experiments on Archimedean tilling pinning arrays 
in superconductors as well as for colloids interacting with optical trap 
arrays with similar geometries.

\section{Acknowledgments}
This work was carried out under the auspices of the 
NNSA of the 
U.S. DoE
at 
LANL
under Contract No.
DE-AC52-06NA25396.

\end{document}